\documentclass[twocolumn,amsmath,aps,superscriptaddress,showpacs,floatfix,a4paper]{revtex4}
\usepackage{amsmath}
\usepackage{graphicx}
\usepackage{epsf}
\usepackage{bm}
\usepackage{latexsym}
\RequirePackage{amsopn}
\RequirePackage{amssymb}
\RequirePackage{amsfonts}
\RequirePackage{amsthm}
\usepackage[]{graphicx}
\usepackage[]{amsmath}

\begin{document}

\title{Coarse-graining dynamics by telescoping down time-scales: comment for Faraday FD144 }

\author{Ard A.\ Louis}
\affiliation{Rudolf Peierls Centre for Theoretical Physics,
           1 Keble Road, Oxford OX1 3NP, United Kingdom}
\date{\today}
\begin{abstract} I briefly review some concepts related to coarse-graining methods for the dynamics of soft matter systems and argue that such schemes will almost always need to telescope down the physical hierarchy of time-scales to a more compressed, but more computationally manageable, separation.
\end{abstract}
\maketitle

The question of how to properly coarse-grain a dynamical simulation is a very interesting one.  I think there is no single answer, but want to use an example here from our simulations of colloidal hydrodynamics to make some points that I believe are of more general relevance.

Consider a buoyant colloid of mass $M_c$ and a radius $a =1 \mu m$ in $H_2O$.   As described in more detail in ~\cite{Padd06},  its behaviour is governed by a series of different timescales shown in  table~\ref{table:time}.   If you are only interested in the behaviour of the colloids, then the two fastest time-scales, the solvent collision time $\tau_{\mathrm{col}}$ and the solvent relaxation time $\tau_f$, can be ignored as long as they are shorter than any other colloidal time-scales.   The first physically relevant time-scale is the Fokker Planck time-scale $\tau_{FP} \approx 10^{-13}$ over which the colloid loses memory of the short-time forces acting on it~\cite{Dhon96}.  For the example colloid, the next  time-scale up is the sonic time $t_{cs} \approx 6.7 \times 10^{-10}s$.   Then comes the Langevin time $\tau_B \approx  2.2 \times 10^{-7} s$ that measures the exponential decay time of the velocity autocorrelation function within the Langevin approximation.  Interestingly, for colloids this time-scale is artificial and does not have direct physical meaning (see appendix of ~\cite{Padd06}).   Next up is the 
 kinematic time $\tau_{\nu}  \approx 10^{-6} s$ over which vorticity diffuses away from the colloid.  If your colloid moves a significant fraction of its radius within the time $\tau_{\nu}$, then the colloid will feel the effects of its own motion from a time $\tau_{\nu}$ back, and finite Reynolds number ($Re$) effects start to kick in.   For that reason, it needs to be kept small compared to  time-scales of colloidal diffusion or advection.   The largest time-scale we consider here is the diffusion time $\tau_D \approx 5s$.  However, if the colloid also moves under an external force with a velocity $v_s$, then there is an additional time-scale $t_s = a/v_s$ that measures how long it takes to  advect over its radius, and you can then also define a related Peclet number $Pe = t_s/\tau_D$ that measures the relative importance of convection over diffusion.  
 
 \begin{figure}[t]
   \scalebox{0.4}{\includegraphics{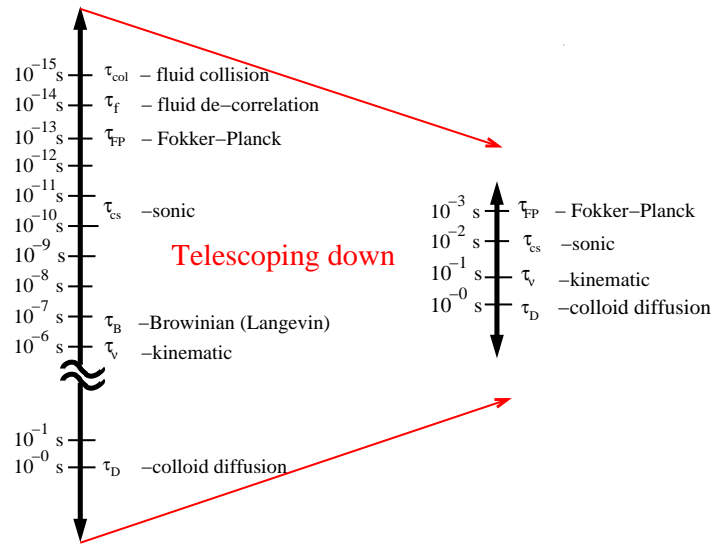}}
\caption{ Telescoping down: The hierarchy of time-scales for a
  colloid (here the example taken is for a colloid of radius 1 $\mu m$
  in H$_2$0) is compressed in the coarse-grained simulations to a more
  manageable separation.  As long as the physically important times are clearly separated, the
  simulation should still generate the correct physical picture.  Once
  the simulations are completed, they can be related in more detail to
  particular experiments by telescoping back out to the relevant
  experimental time-scales.
\label{fig:telescope}
}
\end{figure}
 
 From the Fokker Planck time on up to the diffusion time covers 13 orders of magnitude.   It is clearly not possible to capture all of these in a simulation.
Instead, what is needed is time-scale separation.  As long as the time-scales are properly separated, you should still be simulating the correct underlying physics.  This process can visualized in 
 Figure~\ref{fig:telescope}  (taken from ref.~\cite{Padd06}) which shows an example of how the hierarchy of time-scales is telescoped down to a more computationally manageable separation in order  maximise simulation efficiency, but in such a way that the times are still sufficiently separated to correctly resolve the underlying physical behaviour.   A good example of the rationale behind this thinking can be illustrated with the sonic time $t_{cs}$.  Physically it needs to be much smaller than the diffusion time, or else locally you have supersonic behaviour.  But if it is too small, the simulation will spend most of its time resolving sound waves that may not be that interesting for the colloidal behaviour you are trying to reproduce, making the simulation very inefficient.
 
 In order to correctly interpret the physical meaning of you simulation you need to telescope the time-scale hierarchy back out to the physical one you want to study.        For example, if you are interested in physics that is dominated by diffusion, you would map onto your  physical diffusion time.  A consequence of this strategy is  that a single simulation can map onto many different physical times.  E.g. if your colloid has a radius $a=1 \mu m$, then the diffusion time is $\tau_D = 5 s$, and that fixes the time-scales for your simulation.   On the other hand, if your colloid has a radius $a= 100 nm$, then $\tau_D = 5 \times 10^{-3}s$ and the exact same simulation would have a different fundamental time-scale.      That means that for example the viscosity in your simulations would be different depending on what colloid size you were mapping to, even though the fundamental physics is the same.  This fact suggests that physical viscosity is often not such a good parameter to try and  fit to in a coarse-grained simulation.
 
 You can also correctly map the same simulation to different time-scales even though the colloidal system is unchanged.  Say that you are interested  in the longer-time behaviour of the velocity autocorrelation function that is dominated by the kinematic time $\tau_\nu$.  In that case if you mapped to a physical system with $a = 1 \mu m$, the times would be quite different from what you would get when you mapped to  the diffusion time of the same system.   So here the viscosities etc... would have different values depending on what you processes were focusing on.   This is a good example of a no free lunch theorem.  If you coarse-grain dynamics, you almost always need to do some kind of telescoping down, and that means that it is hard to simultaneously match multiple time-scales in your system.   
 
  The particular example described here
 concerns a colloid in suspension where it is relatively straightforward to work out what all the time-scales are.  Nevertheless, we argue in ~\cite{Padd06}  that many other
 coarse-graining methods for dynamics must  make implicit use of the
 telescoping down process.   You can make your method work by carefully analyzing your time-scales, and then making sure you know how to telescope back out to the experimental situation you want to emulate.   It helps to do this in terms of dimensionless variables.  However sometimes you can't compress the hierarchy to a computationally achievable regime without bringing some time-scales too close to each other, or even  switching the order of time-scales.  It is then a matter of subtle judgement if the behaviour that comes out of your simulation is physically correct.
 
 A similar analysis can be used to interpret systems where the dynamics are dominated by energy barriers. It is going to be very hard to know what the real physical time-scales are here  because processes depend exponentially on barrier heights.  In a coarse-grained dynamical simulation, it may be advantageous to dramatically lower (free) energy barriers in order to speed up the simulation.  The hope is that you keep the relative order that characterizes the physical system you want to emulate, so that at least the qualitative dynamic effects are correct.      But if, say, you try to interpret your time-scales from a measurement of single particle diffusion, and then use that time to extract a physical time for a different process in your simulation that is dominated by energy barriers, then you will get those numbers completely wrong.    Clearly, there are  many tricky subtleties that arise when trying to coarse-grain the dynamics of soft-matter systems.

\begin{table*}
\begin{center}
\caption{Time-scales  relevant for colloidal suspension; numerical  values are for a buoyant colloid of radius $a=1 \mu m$ in $H_2O$.  For $H_2O$, the speed of sound is $1.48 \times 10^9 \mu m/s$. The kinematic viscosity $\nu = 1 \times 10^6 \mu m^2/s$, and measures the diffusion constant with which vorticity diffuses away.  
\label{table:time}
}
\begin{ruledtabular}
\begin{tabular}{lll}

\begin{tabular}{c}
Solvent time-scales   
   \\ \hline \noalign{\medskip}
  \end{tabular}

 \\  

{\it  Solvent collision  time} over which solvent molecules 
interact & $\displaystyle \tau_{\mathrm{col}}  \approx 10^{-15}$~s 
  \\ \noalign{\medskip} 

{\it  Solvent relaxation time} over which solvent velocity correlations
decay & $\displaystyle \tau_f \approx 10^{-14}- 10^{-13}$~s  \\  
\noalign{\medskip} \hline

 \begin{tabular}{c}
Hydrodynamic time-scales   
   \\ \hline \noalign{\medskip}
  \end{tabular}

 \\  

{\it Sonic time} over which sound propagates one colloidal radius &
$\displaystyle t_{cs} = \frac{a}{c_s} \approx 6.7 \times 10^{-10} s$ 
 \\  \noalign{\medskip}

{\it Kinematic time} over which momentum (vorticity) diffuses one colloidal radius  &
$\displaystyle \tau_{\nu} = \frac{a^2}{\nu} \approx 10^{-6} s $ 
  \\  \noalign{\medskip}

\hline

\begin{tabular}{c}
Brownian time-scales   
   \\ \hline \noalign{\medskip}
  \end{tabular}
\\

{\it Fokker-Planck time} over which force-force correlations decay &
$\displaystyle \tau_{FP} \approx  10^{-13} s$  \\  \noalign{\medskip}

{\em Brownian relaxation time} over which colloid velocity correlations decay
in 
 Langevin Eq. & $\displaystyle \tau_B = \frac{M_c}{\xi_S} \approx 2.2 \times 10^{-7} s $  \\  \noalign{\medskip}

 {\it Colloid diffusion time} over which a colloid diffuses over its radius
  & $\displaystyle \tau_D = \frac{a^2}{D_{\mathrm{col}}} \approx 5 s$  \\ \noalign{\medskip} \hline

 \begin{tabular}{c}
  Ordering of time-scales for colloidal particles  \\ \hline 
\end{tabular}  \\ \noalign{\medskip}

  $\displaystyle \tau_{\mathrm{col}} \,\,\,\,\, < \,\,\,\,\,  \tau_{f} , \tau_{FP}\,\,\,\,\ < \,\,\,\,\,\,t_{cs} \,\,\,\,\, < \,\,\,\,\,  \tau_{B}  \,\,\,\,\, <  \,\,\,\,\, \tau_\nu  \,\,\,\,\,  <  \,\,\,\,\, \tau_D $ 

\end{tabular}
\end{ruledtabular}

\end{center}
\end{table*}

\end{document}